\documentclass{PoS}

\usepackage{epsf}
\usepackage{amssymb}
\usepackage{amsmath}
\usepackage{amsfonts}
\usepackage{cite}
\usepackage[small]{caption2}

\usepackage[T1]{fontenc}
\usepackage{psfrag,epsfig,graphicx,graphics}

\newcommand{\cor}[1]{\left\langle{#1}\right\rangle}

\def\eps{\epsilon}

\newcommand{\f}{\frac}
\newcommand{\be}{\begin{equation}}
\newcommand{\beq}{\begin{equation}}
\newcommand{\ee}{\end{equation}}
\newcommand{\eq}{\end{equation}}
\newcommand{\eeq}{\end{equation}}

\newcommand{\rsq}{{\mathfrak R^2}}

\newcommand{\bea}{\begin{eqnarray}}
\newcommand{\eea}{\end{eqnarray}}

\def\slashchar#1{\setbox0=\hbox{$#1$}
   \dimen0=\wd0
   \setbox1=\hbox{/} \dimen1=\wd1
   \ifdim\dimen0>\dimen1
      \rlap{\hbox to \dimen0{\hfil/\hfil}}
      #1
   \else
      \rlap{\hbox to \dimen1{\hfil$#1$\hfil}}
      /gdatdafinal2.tex
   \fi}

\newcommand{\nn}{{\cal N}}
\def\ba{\begin{eqnarray}}
\def\ea{\end{eqnarray}} 
 \def\bi{\begin{itemize}}
 \def\ei{\end{itemize}}
 \def\ii{\item}
\newcommand{\la}{\label}

\title{Bjorken Flow of the quark-gluon plasma and 
 Gauge/Gravity Correspondence}

\ShortTitle{Bjorken Flow of the quark-gluon plasma and 
 Gauge/Gravity Correspondence}

\author{R.~A.~Janik\\
M.Smoluchowski Institute of Physics, Jagellonian University,\\
Reymonta 4, 30-059 Krakow, Poland\\
E-mail: \email{ufrjanik@th.if.uj.edu.pl}}

\author{\speaker{Robi~Peschanski}\\
Institut de Physique Th{\'e}orique and URA 2306, Unit\'{e} de Recherche associ{\'e}e au CNRS, CEA-Saclay, F-91191 Gif/Yvette Cedex, France\\
E-mail: \email{robi.peschanski@cea.fr}}

\abstract{The contribution presents a brief summary of the Gauge/Gravity approach 
to the study of hydrodynamic flow of the quark-gluon plasma formed in heavy-ion collisions,
 in a boost-invariant setting (Bjorken flow). Considering the ideal case of a supersymmetric
  Yang-Mills theory for which the AdS/CFT correspondence gives a precise form of the
   Gauge/Gravity duality, the properties of the strongly coupled expanding plasma are put 
   in one-to-one correspondence with the metric of a 5-dimensional black hole with the horizon moving away
    in the 5th dimension and its deformations consistent with the relevant Einstein 
    equations. Several recently studied aspects of this framework are recalled and put in 
    perspective. New results in collaboration with G.~Beuf and M.~Heller on 
    the early time expansion towards the hydrodynamical regime are provided giving a new 
    insight on the far-from-equilibrium behaviour of the fluid at strong coupling and 
   the thermalization and isotropization problems.
}

\FullConference{European Physical Society Europhysics Conference on High Energy Physics,
EPS-HEP 2009,\\
		 July 16 - 22 2009\\
		 Krakow, Poland}

\begin{document}

\section{Gauge/Gravity 
correspondence and AdS/CFT}
\label{Sec_Int}

The correspondence between Gauge field theories and Gravity has become a general 
tool for investigating problems where QCD perturbative calculations are not expected to work and
 strong coupling seems required. Among those problems, the hydrodynamic behaviour of the Quark-Gluon Plasma
 (QGP) observed in heavy-ion reactions at RHIC is one of the most prominent. Our goal is to 
 use the Gauge/Gravity duality in this context.
 
 Gauge/Gravity duality can be very qualitatively understood in the following intuitive way: 
 in the framework of 10-dimensional string theory, the exchange of a graviton  is 
 equivalent to the one-loop contribution of gauge bosons, since they both are represented by the same
 cylindrical surface. Moreover, while the  large distance (long cylinder)  contribution is particularly simple in the gravity
 language, since it amounts to a one-graviton exchange, it is quite involved in gauge theory,
  since it deals with a quantum one-loop, long open-string, large-distance exchange.
 
 However, for more quantitative study, it is necessary to give a precise meaning to this relationship
 and we are led to consider the paradigmatic case of the Anti de Sitter/$\nn =4$ Supersymmetric
 Yang-Mills theory duality, or in short the AdS/CFT correspondence \cite{Mald}. This duality involves a conformal field theory
 (CFT) and thus has no typical scale, contrary to QCD, where $e.g.\ \Lambda_{QCD}$ plays the role of a confinement scale.
 However, one may expect that lessons from the $\nn =4$ Supersymmetric
 Yang-Mills theory may be relevant for the deconfined phase of a QCD plasma produced in a high-energy heavy-ion collision.
One has to be careful, however, about which observables can be discussed in this way. The bulk properties of a QGP
 are among those where the AdS/CFT correspondence gives some new insight, as we shall see now.

\section{Late Time Dynamics} 
 \label{Sec_Late}

The main practical tool that we will use is the so-called {\it holographic renormalization} \cite{Sken}.
It allows one to relate the properties of the 4-dimensional Minkowskian boundary representing the physical space-time to
properties of gravity in the 5-dimensional AdS space (the 5 other dimensions of the string theory 
play a fundamental role for the consistency of the scheme but not in our calculations). One writes 
\begin{eqnarray}
\label{holography}
ds^2=\f{
g_{\mu\nu}(z)\ dx^\mu dx^\nu+dz^2}{z^2}\quad
g_{\mu\nu}=g^{(0)}_{\mu\nu} 
{(=\eta_{\mu\nu})}+z^2 
g^{(2)}_{\mu\nu} 
{(=0)}+z^4 
{\cor{T_{\mu\nu}}} {~~ + 
z^6{\ldots}} \, .
\end{eqnarray}
In (\ref{holography}), $g_{\mu\nu}(z)$ is  the Fefferman-Graham (FG)
 5-dimensional metric, $\cor{T_{\mu\nu}}$ being the expectation value of the physical 4-dimensional
 stress-energy tensor.  The {\it holography} property, that is the one-to-one correspondence between the 4d-boundary
  and the 5d-bulk is here realized by the property that all subsequent coefficients $
z^6{\ldots}$ of 
  the  $z$-expansion in the fifth dimension, are determined by $\cor{T_{\mu\nu}}$ $via$ the 
  5-dimensional Einstein equations, provided the stress-energy tensor verifies the tracelessness and energy-momentum conservation relations. 

The main goal of our approach  \cite{Jani} is to derive the gravity duals of an expanding fluid corresponding
to the plasma phase of the $\nn =4$ Supersymmetric
 Yang-Mills theory in the relativistic kinematics proper to a heavy-ion reaction. In this 
 section we recall previous results of our approach concerning the {\it late-time} properties 
 of a boost-invariant  flow. Indeed, the complexity of the Einstein equations to be solved is reduced by 
 the boost-invariant symmetry and the existence of a hierarchy between the coefficients of the expansion \eqref{holography}
due to a scaling property.
 Starting with a family of stress-energy tensors
 \begin{eqnarray}
\label{tensors}
\cor{T_{\mu\nu}}={\rm Diag}\ \left\{\eps(\tau),\f{d}{d\tau} 
\eps(\tau)\!-\!\tau^2 \eps(\tau),\eps(\tau)\!+\! \f{1}{2}\tau\f{d}{d\tau} 
\eps(\tau),
\eps(\tau)\!+\! \f{1}{2}\tau\f{d}{d\tau} \eps(\tau)\right\} ,
\end{eqnarray}
expressed in the $\tau-\eta$ proper-time-spatial rapidity frame, 
with $\eps(\tau) \propto \tau^{-s}$ 
and 
applying the holographic renormalization by resumming Eq.\eqref{holography},
one realizes that the only consistent (i.e. nonsingular) solution is given by 
$s\equiv \f 43,$
which corresponds to the perfect fluid (or Bjorken flow). Moreover for that value, the 
obtained FG metric is a Black Hole one (more precisely a black brane)
with the horizon moving away in the $z$-direction, in a holographic correspondence with the falling temperature due to cooling
of  the plasma. Indeed, precise duality relations exist at large $\tau$
between the moving Black Hole and the cooling Plasma, namely
\begin{eqnarray}
{Horizon:}\ 
z_h\propto \tau^{\f{1}{3}}   
\quad {Temperature:} \sim  
\tau^{-\f{1}{3}}
\quad {Entropy}\ \sim Area:  
\sim \tau \cdot {1}/{z_h^3} =
const
\label{BH}\, .
\end{eqnarray}
One recognizes here the properties of an ideal Bjorken flow.

Taking into account the asymptotic limit of the perfect fluid, it has been
possible to derive an expansion at large $\tau,$ allowing to derive the 
dissipative corrections of the flow due to dissipation and transport 
through the dual gravitational properties. The main ingredients are 
the same, $i.e.$ the expansion of Einstein equations and the consistency 
of the holographic renormalization requiring nonsingular geometries. For instance , 
beyond the perfect fluid, one can compute the shear viscosity through the equation
$\partial_{\tau}\epsilon= -\f 43 \f{\epsilon}{\tau }+\f{\eta}{\tau^{2}}+\cdots
\Rightarrow {\f \eta s = \f 1{4\pi}},$ which gives the same result
as the classical static calculations. 

\section{Early Time Dynamics}
\la{Sec-Early}

The problem of early-time dynamics, even with the boost-invariance symmetry assumption, appears to be more involved. In fact one is faced with
the full set of nonlinear Einstein equations 
\ba
R_{AB} - \frac{1}{2} R \cdot G_{AB} - 6 \, G_{AB} = 0\mathrm{,}
\la{einstein}
\ea
where $R_{AB}$ is the Ricci tensor  and  the FG metric $G_{AB}$ is given by
\ba
ds^2= G_{AB}dx_A dx_B =\f{-e^{a(\tau,z)}\ d\tau^2 +\tau^2 e^{b(\tau,z)}\  dy^2
  +e^{c(\tau,z)}\  dx^2_\perp}{z^2} +\f{dz^2}{z^2}\ .
\la{general}
\ea 
Indeed, one is physically dealing with a non-(even far-from) equilibrium fluid whose
theoretical properties and formalism are hardly known. It is a remarkable but challenging
feature 
 of  Gauge/Gravity duality to give first hints on this difficult problem. In a very recent work \cite{Beuf},
 we have obtained some results following the same path as for late time (holography and nonsingularity
  of the geometry) but with quite different features from the late-time solution.
  
  \bi
  \ii 
  {\it No scaling at early time.} In fact, there appears to be no hierarchy 
  at small $\tau$ between the 
  coefficients of the holographic expansion \eqref{holography}. The full content of the Einstein equations has
   to be taken into account. 
  \ii {\it Dependence on initial conditions.} Due to the absence of scaling, there is no
unique geometry analogous to the moving Black Hole  solution of late time.  In physical terms, the far-from equilibrium 
  pre-asymptotic regime of the flow is mainly governed by the initial conditions, and in the
   dual set-up
   by the gravitational response to them.
   \ii {\it The metric is singular at all times.} Quite unexpectedly, it is possible to derive a general 
   mathematical consequence of the Einstein equations (essentially from the $G_{zz}$
    component) following which there should be a singularity in $z$ 
     at all proper-time  $\tau$
    in the metric 
    \eqref{general}. Hence some kind of non-trivial horizon exists from the beginning of the flow, which
    can be interpreted as the ``ancestor'' of the BH horizon of the late-time solution.
   \ii{\it The geometry should stay regular.} Following the same consistency requirement as previously,
   one is led to impose the constraint of a regular geometry, meaning that the singularity in the 
   metric is only a coordinate singularity with the curvature scalars
    (such as $\rsq$) remaining bounded analogously as for a Schwarzschild horizon.
  \ei
  
\begin{figure}
  \centering
  \includegraphics[width=5cm]{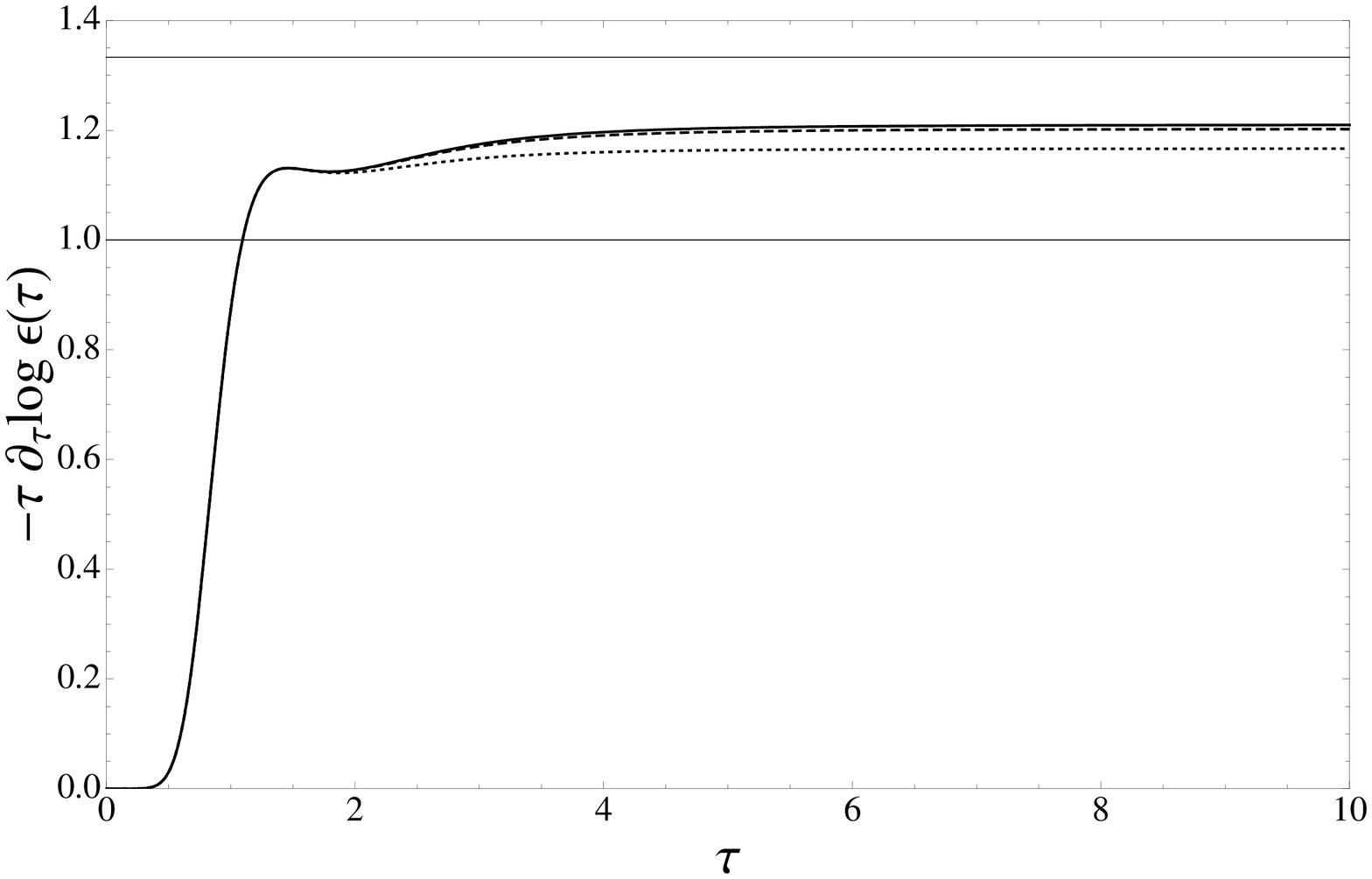}\hspace{10mm}
  \includegraphics[width=5cm]{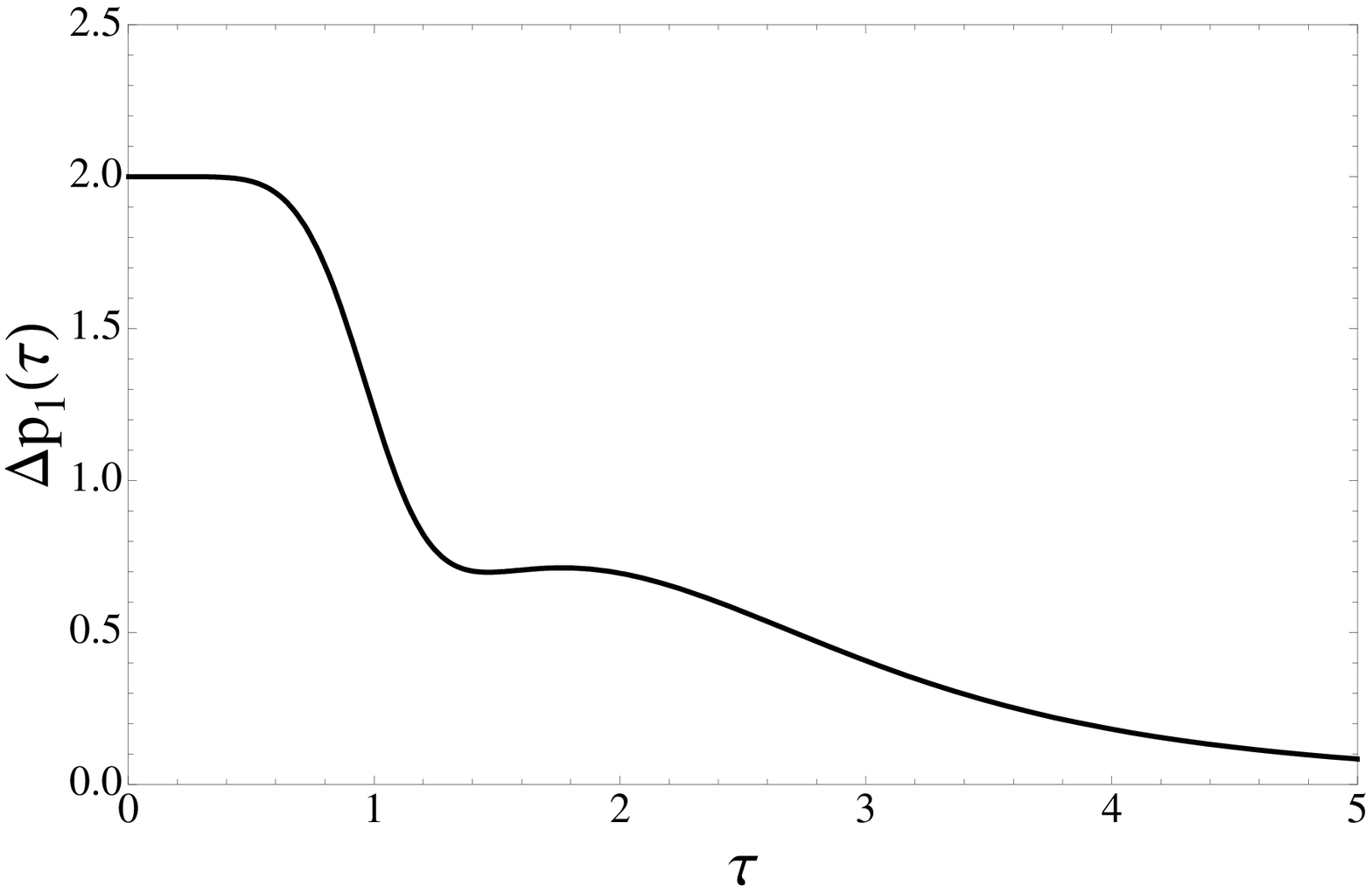}
  \caption{Left: The index $s$ from Energy density;  Right: Pressure anisotropy. 
}
\label{1}
\end{figure}

  In practice, one method is to solve the Einstein equations 
\eqref{einstein} in a power series
  with an appropriate ansatz for the initial conditions $e.g.$ for the coefficient $a(z,\tau=0)$
   of the metric \eqref{general}. Among the first obtained results, we show in fig.~\ref{1},
the result for the index $s=\frac {d\log \eps(\tau)}{d\tau}$  and the pressure
 anisotropy $\Delta p \left( \tau \right) = 1 - \frac{p_{\parallel} 
\left( \tau \right)}{p_{\perp} \left( \tau \right)}$ for one typical solution.  A quite general
 feature seems to emerge consisting in  a quite fast evolution to thermalization
 at the beginning but  
with eventually a partial isotropization, as shown in fig.~\ref{1}. However, 
since it was necessary to perform a resummation procedure, 
before a firm conclusion is reached, a complete study
using direct numerical solution of Einstein's equations has to be made.

{\bf Acknowledgements:} RJ was supported by Polish science funds during 2009-2011 as a research project (NN202 105136).

\end{document}